\documentclass[epj]{webofc}\usepackage[varg]{txfonts}\usepackage{bm}
\woctitle{QCD@Work 2014}\begin{document}\title{Charmed
Pseudoscalar and Vector Mesons: a Comprehensive\\QCD Sum-Rule View
of Their Decay Constants}\author{Wolfgang Lucha\inst{1}\fnsep
\thanks{\email{Wolfgang.Lucha@oeaw.ac.at}}\and Dmitri Melikhov
\inst{2,3}\fnsep\thanks{\email{dmitri_melikhov@gmx.de}}\and
Silvano Simula\inst{4}\fnsep\thanks{\email{simula@roma3.infn.it}}}
\institute{Institute for High Energy Physics, Austrian Academy of
Sciences, Nikolsdorfergasse 18, A-1050 Vienna, Austria\and Faculty
of Physics, University of Vienna, Boltzmanngasse 5, A-1090 Vienna,
Austria\and D.~V.~Skobeltsyn Institute of Nuclear Physics,
M.~V.~Lomonosov Moscow State University, 119991, Moscow,
Russia\and INFN, Sezione di Roma Tre, Via della Vasca Navale 84,
I-00146, Roma, Italy}\abstract{In spite of undeniable similarities
of the applied techniques, somewhat different challenges are
encountered when extracting, from QCD sum rules derived from
two-point correlators of appropriate interpolating heavy--light
quark currents, the decay constants of charmed mesons of
\emph{pseudoscalar\/} nature, on the one hand, or of
\emph{vector\/} nature, on the~other hand. This observation
justifies a rather careful reassessment of the corresponding
results.}\maketitle

\section{Introduction: QCD sum rules in a nutshell---techniques and
applications}\label{Sec:QSR}\emph{QCD sum rules\/} \cite{SVZ} are
relations between features of hadrons---the bound states governed
by the strong interactions---and the parameters of their
underlying quantum field theory---QCD. Such relations may be
established (rather straightforwardly) by analyzing vacuum
expectation values of nonlocal products of interpolating
operators---specifically, of appropriate quark currents---at both
QCD and hadron level. Upon application of Wilson's \emph{operator
product expansion\/} (OPE) for casting, at QCD level, any arising
nonlocal operator product into the form of a series of local
operators, contributions of both perturbative as well as
nonperturbative (NP) origin enter: the former are usually
represented by dispersion integrals of certain spectral densities
while the latter---also called the ``power'' corrections---involve
the vacuum expectation values of all local OPE operators---crucial
quantities going, in this context, under the name of ``vacuum
condensates.'' Then, performing a \emph{Borel transformation\/}
from one's momentum~variable to a new variable, the Borel
parameter $\tau,$ lessens the importance of hadronic excited and
continuum~states for such ``Borelized'' sum rules and removes
potential subtraction terms. Our lack of knowledge about higher
states is dealt with by postulating \emph{quark--hadron
duality\/}: all contributions of hadron excited and continuum
states roughly cancel against those of perturbative QCD above an
effective threshold~$s_{\rm eff}(\tau).$

Here, after sketching, in Sect.~\ref{Sec:DCCM}, the QCD sum-rule
extraction of heavy-meson decay constants and recalling, in
Sect.~\ref{Sec:AEHF}, a few ideas for improvement of this concept,
we focus to its intrinsic uncertainties. Its systematic errors are
subject to at least two effects demanding our attention: an
\emph{optimal\/} perturbative behaviour, discussed in
Sect.~\ref{Sec:IMPC}, and the \emph{fake\/} impact of the
renormalization scale, $\mu,$ analyzed in Sect.~\ref{Sec:IRSD}.

\section{\boldmath Charmed pseudoscalar ($D_{(s)}$) and vector
($D^*_{(s)}$) meson decay constants}\label{Sec:DCCM}In order to
predict the decay constants $f_{\rm P,V}$ of charmed pseudoscalar
(P) and vector (V) mesons of mass $M_{\rm P,V}$ considered as
bound states of a charmed quark $c$ of mass $m_c$ and a light
quark $q=d,s$ of mass $m_q,$ we use two-point correlators of
appropriate currents, given in terms of the QCD degrees of
freedom, to find QCD sum rules involving both spectral densities
$\rho^{\rm(P,V)}(s,\mu)$ and nonperturbative terms $\Pi_{\rm
NP}^{\rm(P,V)}(\tau,\mu)$ at the relevant renormalization scale(s)
$\mu.$ Pseudoscalar currents then yield, for \emph{pseudoscalar
mesons\/}~P,\begin{equation}f_{\rm P}^2\,M_{\rm
P}^4\exp\left(-M_{\rm P}^2\,\tau\right)
=\int_{(m_c+m_q)^2}^{s_{\rm eff}(\tau)}{\rm d}s\,{\rm
e}^{-s\,\tau}\rho^{\rm(P)}(s,\mu)+\Pi_{\rm NP}^{\rm(P)}(\tau,\mu)
\equiv\widetilde\Pi_{\rm P}(\tau,s_{\rm eff}(\tau))\
,\label{Eq:PSSR}\end{equation}which suggests to define \emph{dual
masses\/} and \emph{dual decay constants\/} via the \emph{dual
correlator\/} $\widetilde\Pi_{\rm P}(\tau,s_{\rm eff}(\tau))$~by
\begin{equation}M_{\rm dual}^2(\tau)\equiv-\frac{{\rm d}}{{\rm
d}\tau}\log\widetilde\Pi_{\rm P}(\tau,s_{\rm eff}(\tau))\ ,\qquad
f_{\rm dual}^2(\tau)\equiv\frac{\exp\left(M_{\rm
P}^2\,\tau\right)}{M_{\rm P}^4}\,\widetilde\Pi_{\rm P}(\tau,s_{\rm
eff}(\tau))\ .\label{Eq:PSDMDC}\end{equation}Starting, however,
from vector currents yields the counterparts of
Eqs.~(\ref{Eq:PSSR}) and (\ref{Eq:PSDMDC}) for \emph{vector
mesons\/}~V:\begin{align}&f_{\rm V}^2\,M_{\rm
V}^2\exp\left(-M_{\rm V}^2\,\tau\right)
=\int_{(m_c+m_q)^2}^{s_{\rm eff}(\tau)}{\rm d}s\,{\rm
e}^{-s\,\tau}\rho^{\rm(V)}(s,\mu)+\Pi_{\rm
NP}^{\rm(V)}(\tau,\mu)\equiv\widetilde\Pi_{\rm V}(\tau,s_{\rm
eff}(\tau))\ ,\\&\,M_{\rm dual}^2(\tau)\equiv-\frac{{\rm d}}{{\rm
d}\tau}\log\widetilde\Pi_{\rm V}(\tau,s_{\rm eff}(\tau))\ ,\qquad
f_{\rm dual}^2(\tau)\equiv\frac{\exp\left(M_{\rm
V}^2\,\tau\right)}{M_{\rm V}^2}\,\widetilde\Pi_{\rm V}(\tau,s_{\rm
eff}(\tau))\ .\end{align}For our OPE input required at QCD level,
we use the rather standard set of parameter values in
Table~\ref{Tab:NPV}.

\begin{table}[h]\centering\caption{Numerical parameter values
employed as input to the charmed-meson operator product
expansions.}\label{Tab:NPV}\begin{tabular}{cc}\hline\hline&
\\[-1.5ex]Quantity&Numerical input value\\[1ex]\hline&\\[-1.5ex]
$\overline{m}_d(2\;\mbox{GeV})$&$(3.42\pm0.09)\;\mbox{MeV}$\\[.5ex]
$\overline{m}_s(2\;\mbox{GeV})$&$(93.8\pm2.4)\;\mbox{MeV}$\\[.5ex]
$\overline{m}_c(\overline{m}_c)$&$(1275\pm25)\;\mbox{MeV}$\\[.5ex]
$\alpha_{\rm s}(M_Z)$&$0.1184\pm0.0020$\\[1ex]$\langle\bar
qq\rangle(2\;\mbox{GeV})$&$-[(267\pm17)\;\mbox{MeV}]^3$\\[.5ex]
$\langle\bar ss\rangle(2\;\mbox{GeV})$&
$(0.8\pm0.3)\times\langle\bar qq\rangle(2\;\mbox{GeV})$\\[.5ex]
$\displaystyle\left\langle\frac{\alpha_{\rm
s}}{\pi}\,GG\right\rangle$& $(0.024\pm0.012)\;\mbox{GeV}^4$
\\[1.5ex]\hline\hline\end{tabular}\end{table}

\section{Improving QCD sum rules by advanced extraction of hadronic
properties}\label{Sec:AEHF} The \emph{accuracy\/} of QCD sum-rule
predictions for hadronic observables extracted by rather
long-standing traditional techniques \cite{SVZ} may be
significantly improved by dropping the requirement of Borel
stability \cite{LMSAUa,LMSAUb,LMSAUc,LMSAUd,MAU}---reflecting
merely the \emph{prejudice\/} that the value of any such
observable at its extremum in $\tau$ forms a good approximation to
its actual value---and the perhaps very na\"ive \emph{belief\/}
that the effective threshold at QCD level does not know about
$\tau$ \cite{LMSETa,LMSETb,LMSETc,LMSETd,LMSSET}: Earlier analyses
\cite{LMSAUa,LMSAUb,LMSAUc,LMSAUd,MAU} (backed up by quantum
mechanics, where exact solutions may be derived by just solving
Schr\"odinger equations) forced us to conclude~that predictions
relying on Borel stability may emerge rather far from the truth
and that effective thresholds \emph{will\/} depend on $\tau;$ they
culminated in a simple \emph{prescription\/}
\cite{LMSETa,LMSETb,LMSETc,LMSETd,LMSSET} for the extraction of
hadron features:\begin{itemize}\item The \emph{admissible\/}
$\tau$ \emph{range\/} is determined by requiring, at the lower
end, the ground-state contribution to be sufficiently large and,
at the upper end, the power-correction contributions to be
reasonably small. For the charmed pseudoscalar and vector mesons,
the demands may be satisfied if choosing~as Borel windows
$0.1\;\mbox{GeV}^{-2}<\tau<0.5\;\mbox{GeV}^{-2}$ for $D,$ $D^*,$
$D_s^*$ or $0.1\;\mbox{GeV}^{-2}<\tau<0.6\;\mbox{GeV}^{-2}$ for
$D_s$ \cite{LMSDa,LMSDb,LMSD*}.\item The \emph{functional
dependence\/} of the threshold $s_{\rm eff}(\tau)$ on $\tau$ is
modelled by adopting a power-law Ansatz,\begin{equation}
s^{(n)}_{\rm eff}(\tau)=\sum_{j=0}^ns_j\,\tau^j\ ,\label{EQ:PLA}
\end{equation} with expansion coefficients $s_j$ determined by
minimizing, over a set of $N$ equidistant discrete points,
$\tau_i,$ in the allowable $\tau$ range, the deviation of the
predicted from the measured meson masses squared:
\begin{equation}\chi^2\equiv\frac{1}{N}\sum_{i=1}^N\left[M^2_{\rm
dual}(\tau_i)-M_{\rm P,V}^2\right]^2\ .\end{equation}\item
Remembering a lesson drawn from quantum-mechanical analogues of
QCD sum rules, the spread~of results found for the order $n=1,2,3$
of the Ansatz (\ref{EQ:PLA}) is taken as hint of the
\emph{intrinsic sum-rule~error}.\end{itemize}

\begin{figure}[h]\centering\begin{tabular}{cc}
\includegraphics[scale=.5396]{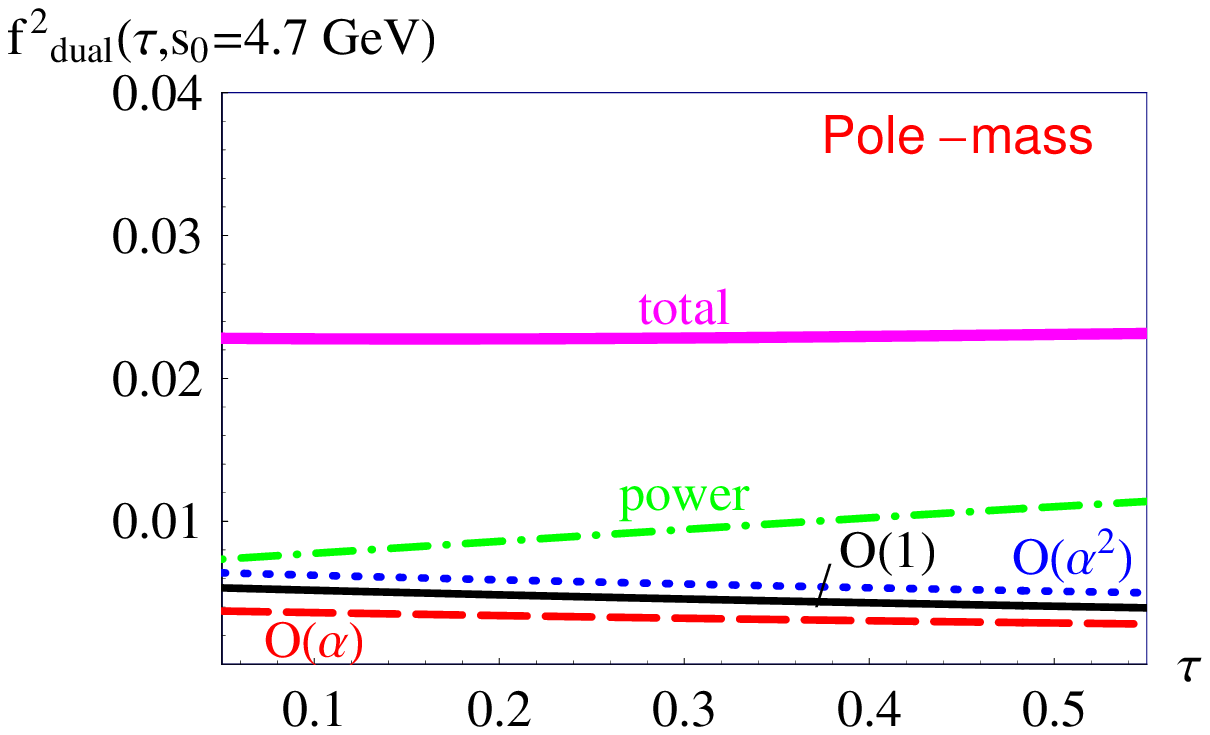}&
\includegraphics[scale=.5396]{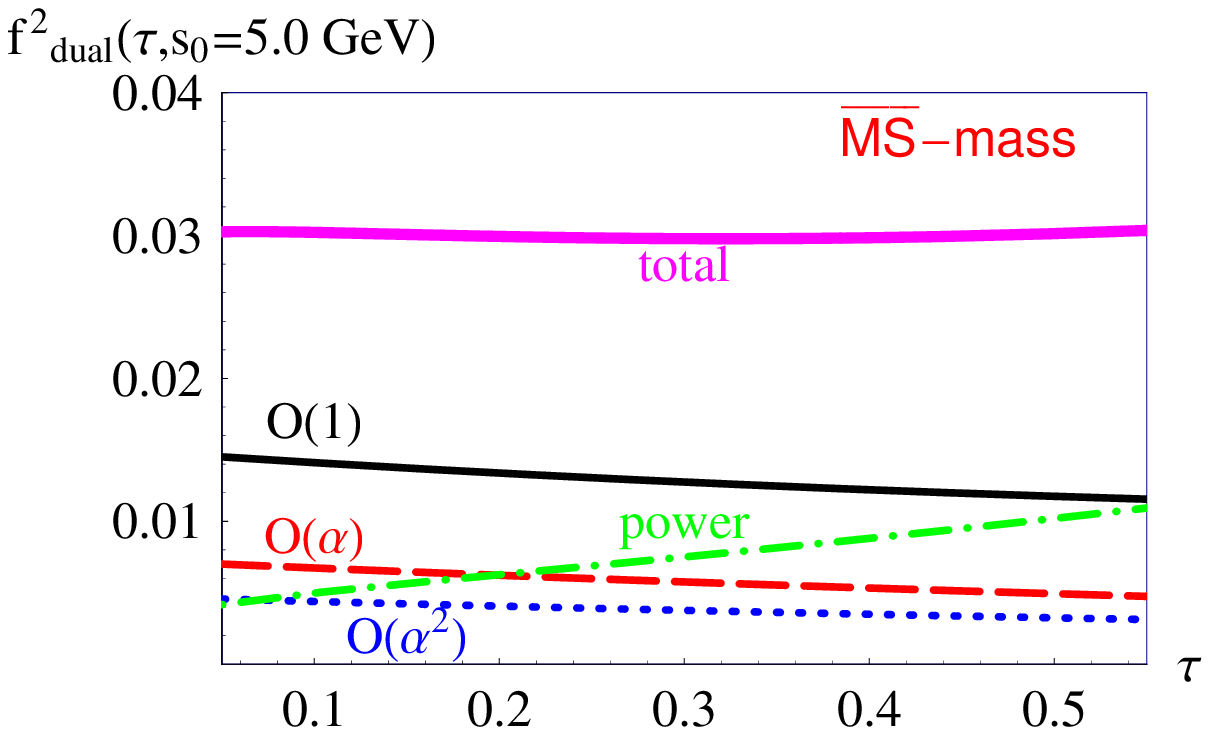}\\
\includegraphics[scale=.412]{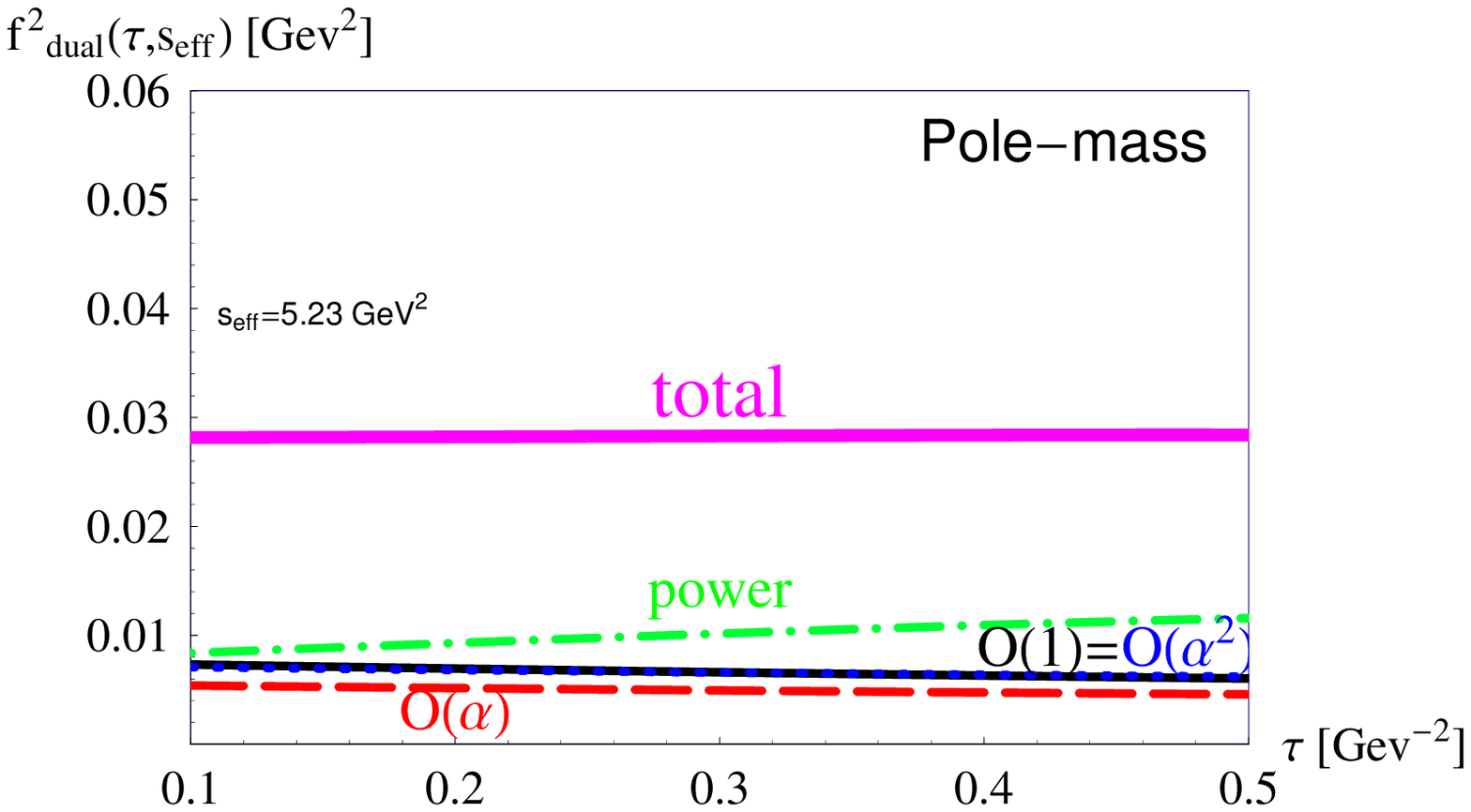}&
\includegraphics[scale=.3932]{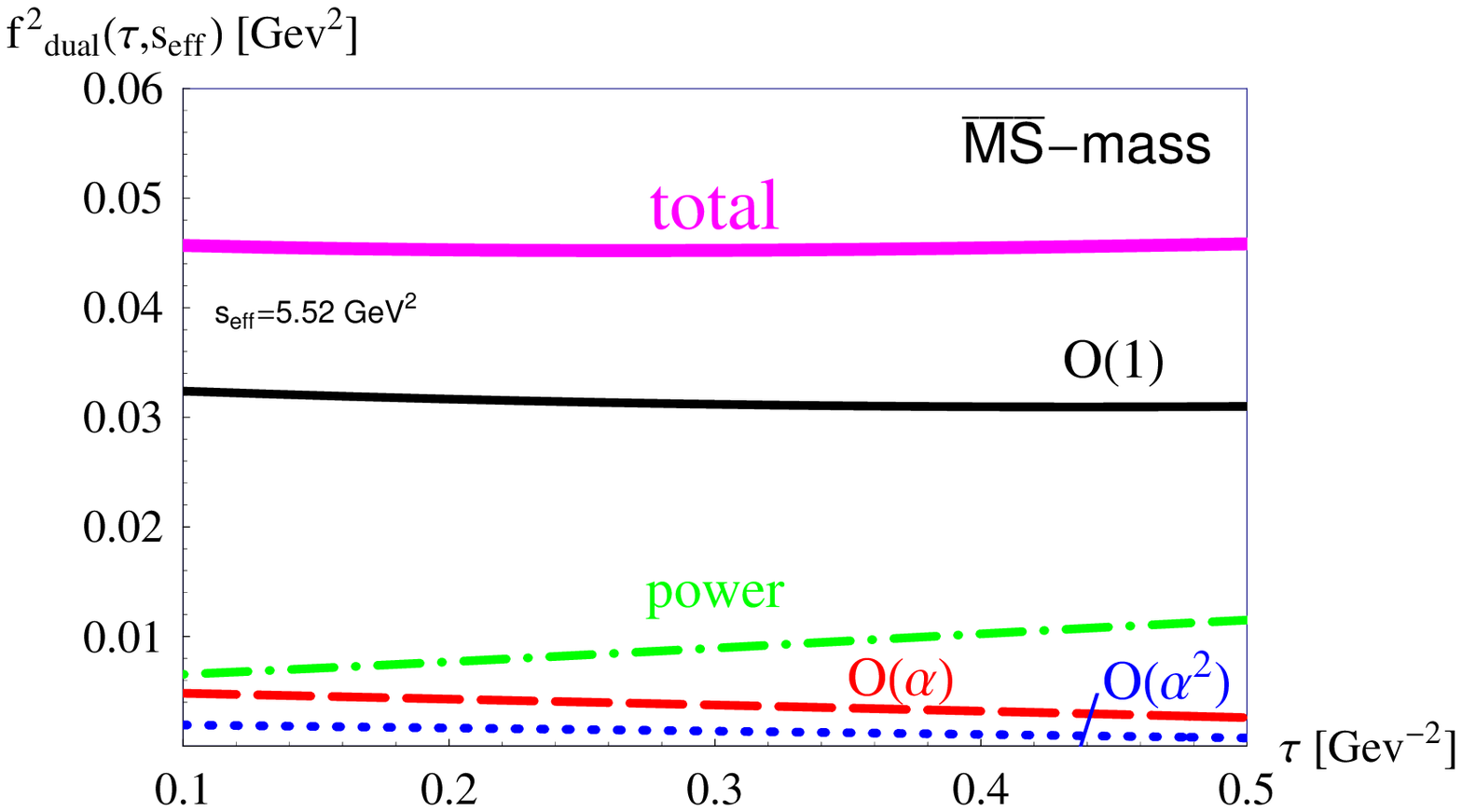}\\
\end{tabular}\caption{Hierarchy of the OPE contributions to the
dual decay constants, $f_{\rm dual}(\tau),$ for both charmed
\emph{pseudoscalar\/} meson $D$ (top row) and charmed
\emph{vector\/} meson $D^*$ (bottom row), as obtained in the
\emph{pole-mass\/} (left column)~and the
\emph{$\overline{\textit{MS}}$-mass\/} (right column)
renormalization scheme, for fixed threshold called $s_0$ in the
$D$ case and $s_{\rm eff}$ in the~$D^*$~case.}\label{Fig:HOPE}
\end{figure}

\section{Topical issue: Maximal perturbative convergence
\cite{LMSDa,LMSDb,LMSD*}}\label{Sec:IMPC}Within \emph{perturbation
theory\/}, each of the coefficients of the various local operators
in the OPE is derived in the shape of a series in powers of the
strong coupling, $\alpha_{\rm s}(\mu).$ In particular, the
coefficient multiplying the unit operator ends up in the spectral
density, presently determined to three-loop $\left(\alpha_{\rm
s}^2\right)$ order \cite{SDa,SDb}:

\begin{equation}\rho(s,m_c,\mu)=\rho_0(s,m_c)+\frac{\alpha_{\rm
s}(\mu)}{\pi}\,\rho_1(s,m_c)+\frac{\alpha_{\rm s}^2(\mu)}{\pi^2}\,
\rho_2(s,m_c,\mu)+\cdots\ .\end{equation}The rate of convergence
of perturbative findings is sensitive to the renormalization
scheme defining the $c$-quark's mass. In this respect, using its
\emph{$\overline{\rm MS}$ running mass\/} $m_c=
\overline{m}_c(\overline{m}_c)=(1275\pm25)\;\mbox{MeV}$~is~superior
to adopting its \emph{pole mass\/} $m_c=\mathring{m}_c=1699\;{\rm
MeV},$ being related by means of given expressions
$r_{1,2}$~\cite{JL}:
\begin{equation}\overline{m}_c(\mu)=\mathring{m}_c\,
\left(1+\frac{\alpha_{\rm s}(\mu)}{\pi}\,r_1+\frac{\alpha_{\rm
s}^2(\mu)}{\pi^2}\,r_2+\cdots\right) .\end{equation}Inspecting
Fig.~\ref{Fig:HOPE}, the gain in perturbative credibility is
evident and visibly larger for the vector mesons.

\section{Topical issue: Renormalization-scale dependence
\cite{LMSDa,LMSDb,LMSD*}}\label{Sec:IRSD}
\begin{figure}[b]\centering\begin{tabular}{cc}
\includegraphics[scale=.42122]{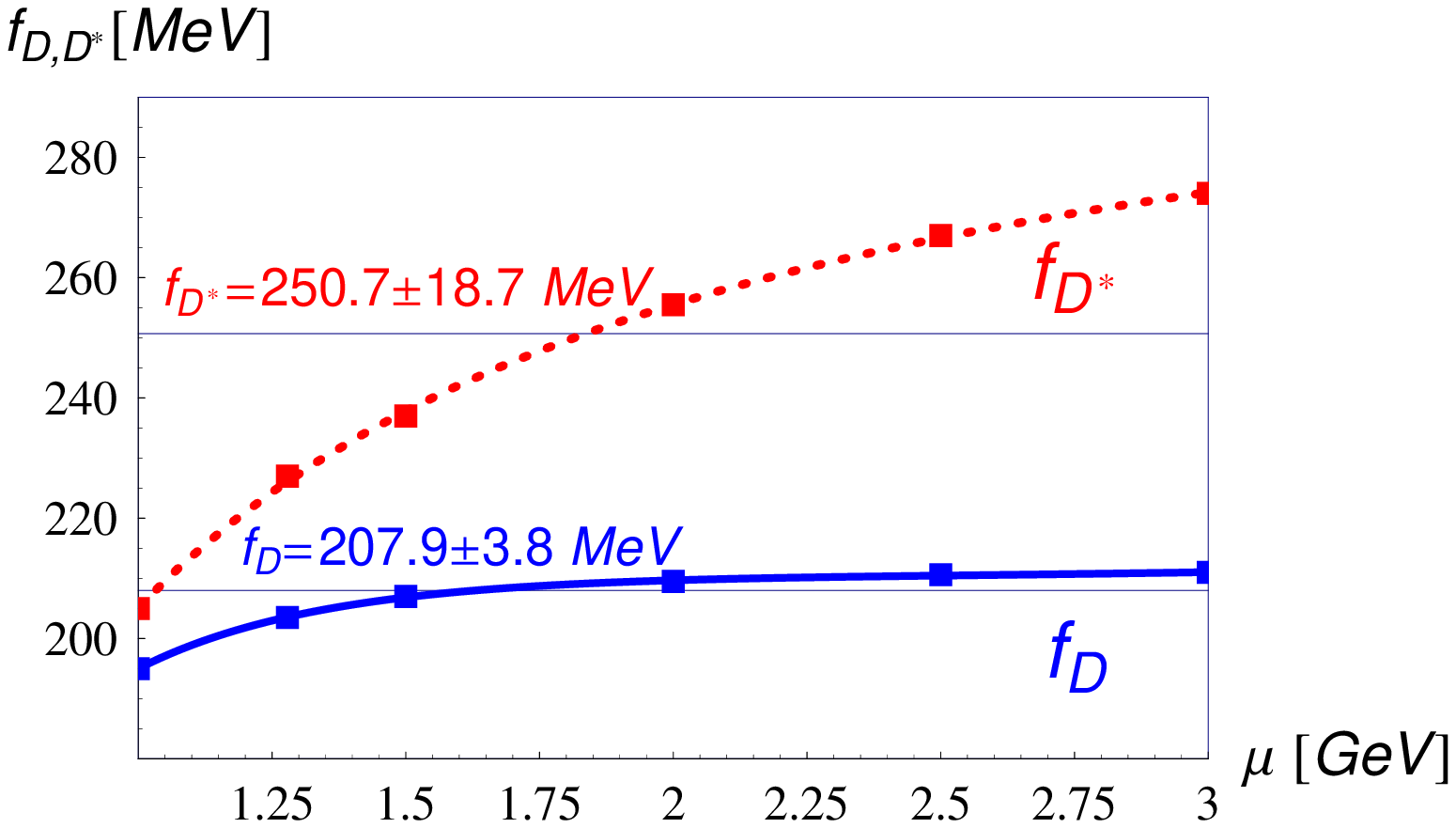}&
\includegraphics[scale=.42122]{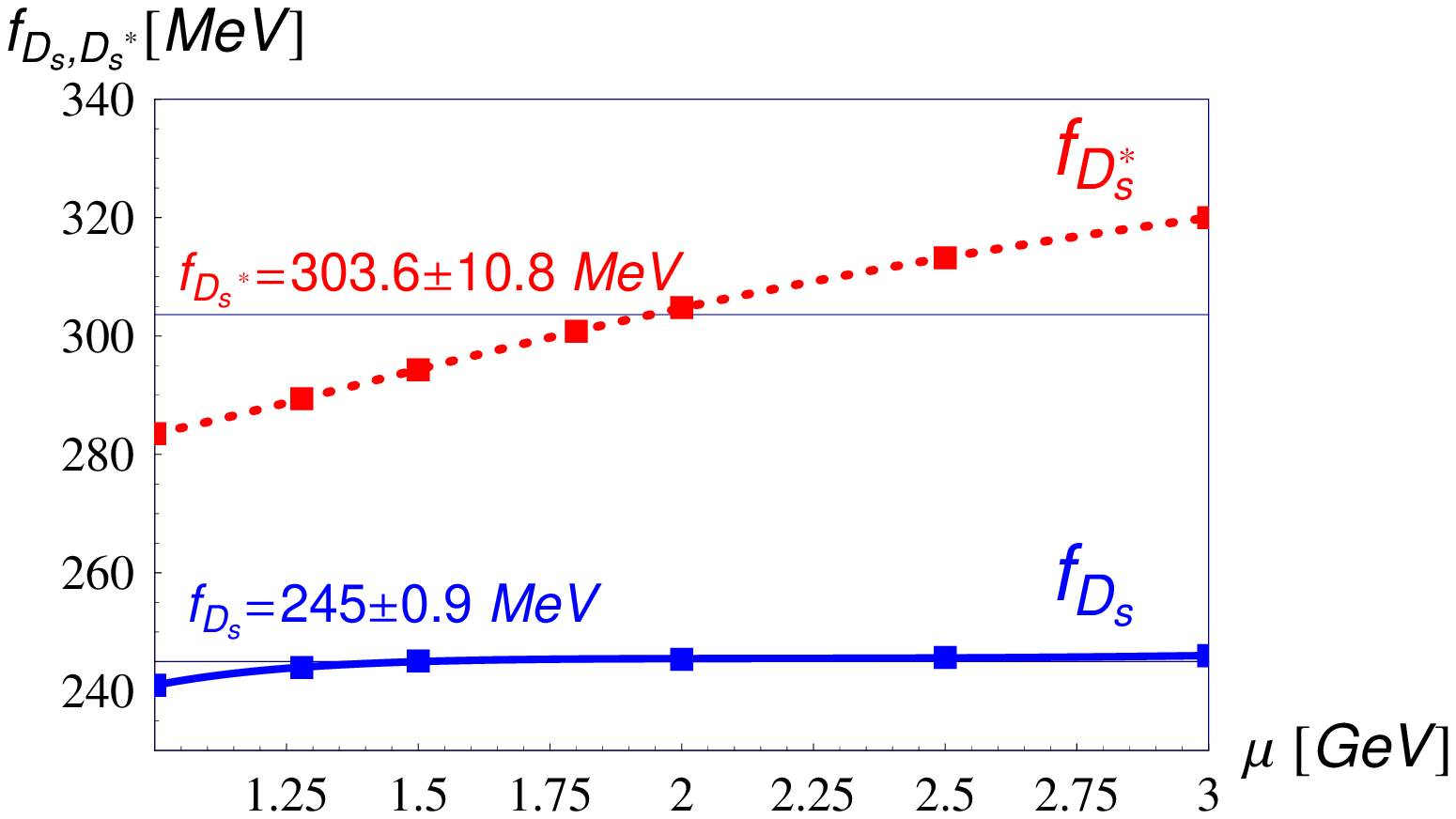}\end{tabular}
\caption{Dependences on the renormalization scale $\mu$ of our QCD
sum-rule findings, for the dual decay constants of the charmed
\emph{non-strange\/} mesons $D$ and $D^*$ ($f_{D^{(*)}}$, left),
and the charmed \emph{strange\/} mesons $D_s$ and $D_s^*$
($f_{D_s^{(*)}}$,~right).}\label{Fig:RSD}\end{figure}

Needless to say, exact
correlation functions do \emph{not\/} depend on any
renormalization scale(s) $\mu.$ However, due to practically
inevitable truncations to finite-order perturbative expansions or
to finite-dimensional vacuum condensates, spectral densities and
power corrections and thus \emph{predicted\/} hadronic features
do. Defining an average $\overline{\mu}$ of the renormalization
scale $\mu$ by requiring $f_{\rm dual}(\overline{\mu})=\langle
f_{\rm dual}(\mu)\rangle,$ such \emph{unphysical\/} decay-constant
sensitivity to $\mu$ is more pronounced for vector than for
pseudoscalar mesons, see Fig.~\ref{Fig:RSD}:\begin{align}
f_D(\mu)&=208.3\;\mbox{MeV}\,\left[1+0.06\log(\mu/\overline{\mu})
-0.11\log^2(\mu/\overline{\mu})+0.08\log^3(\mu/\overline{\mu})\right],\\
f_{D_s}(\mu)&=246.0\;\mbox{MeV}\,\left[1+0.01\log(\mu/\overline{\mu})
-0.03\log^2(\mu/\overline{\mu})+0.04\log^3(\mu/\overline{\mu})\right],\\
f_{D^*}(\mu)&=252.2\;\mbox{MeV}\,\left[1+0.233\log(\mu/\overline{\mu})
-0.096\log^2(\mu/\overline{\mu})+0.17\log^3(\mu/\overline{\mu})\right],\\
f_{D_s^*}(\mu)&=305.5\;\mbox{MeV}\,\left[1+0.124\log(\mu/\overline{\mu})
+0.014\log^2(\mu/\overline{\mu})-0.034\log^3(\mu/\overline{\mu})\right].
\end{align}Table \ref{Tab:ARS} tells us that the averages
$\overline{\mu}$ are somewhat larger for the vector than for the
pseudoscalar mesons.

\begin{table}[ht]\centering\caption{Numerical values of the average
renormalization scales $\overline{\mu}$ for the charmed-meson dual
decay constants.}\label{Tab:ARS}
\begin{tabular}{lllll}\hline\hline&&&&\\[-1.5ex]Meson&
\multicolumn{1}{c}{$D$}&\multicolumn{1}{c}{$D_s$}&
\multicolumn{1}{c}{$D^*$}&\multicolumn{1}{c}{$D_s^*$}\\[1ex]
\hline&&&&\\[-1.5ex]$\overline{\mu}\;\mbox{(GeV)}$
&1.62&1.52&1.84&1.94\\[1ex]\hline\hline\end{tabular}\end{table}

\section{Observations, outcomes, and conclusions}\label{Sec:OO&C}
This \emph{simultaneous\/} scrutiny of QCD sum-rule predictions
for the decay constants of the charmed vector \cite{LMSD*} and
pseudoscalar \cite{LMSDa,LMSDb} mesons discloses, not
surprisingly, both similarities and dissimilarities: With respect
to the perturbative convergence of the extraction procedures, both
types of mesons prefer, beyond doubt, the use of the
$\overline{\textit{MS}}$ definition for the heavy quark's mass.
The effects of this are important for vector and pseudoscalar
mesons. For both types of mesons, the calculated central values
relying on the $\overline{{\rm MS}}$ mass are significantly
\emph{larger\/} than those emerging from the pole mass.
Pseudoscalar mesons~do not seem to care too much about the precise
value of the renormalization scale $\mu,$ whereas its impact on
the \emph{vector\/} mesons is not negligible for their OPE-related
errors. Our results for the decay constants~are
\begin{align}f_D&=\left(206.2\pm7.3_{\rm OPE}\pm5.1_{\rm
syst}\right)\mbox{MeV}\ ,&\quad f_{D_s}&=\left(245.3\pm15.7_{\rm
OPE}\pm4.5_{\rm syst}\right)\mbox{MeV}\
,\\f_{D^*}&=\left(252.2\pm22.3_{\rm OPE}\pm4_{\rm
syst}\right)\mbox{MeV}\ ,&\quad f_{D_s^*}&=\left(305.5\pm26.8_{\rm
OPE}\pm5_{\rm syst}\right)\mbox{MeV}\ .\end{align}

\end{document}